\documentclass{aa}  

\usepackage{graphicx}
\usepackage{natbib}
\bibpunct{(}{)}{;}{a}{}{,} % to follow the A&A style
\usepackage[varg]{txfonts}
\begin{document}

   \title{Variable millimetre radiation from the colliding-wind binary 
Cyg~OB2~\#8A\thanks{This work is based on observations carried out under project numbers 
                    S14AW and S16AU with the IRAM NOEMA Interferometer. IRAM 
                    is supported by INSU/CNRS (France), MPG (Germany) and 
                    IGN (Spain).}
         }

   \subtitle{}

   \author{R. Blomme
          \inst{1}
          \and
          D. M. Fenech\inst{2}
          \and
          R. K. Prinja\inst{2}
          \and
          J. M. Pittard\inst{3}
          \and
          J. C. Morford\inst{2}
          }

   \institute{Royal Observatory of Belgium,
              Ringlaan 3, B-1180 Brussels, Belgium\\
              \email{Ronny.Blomme@oma.be}
         \and
              Department of Physics and Astronomy,
              University College London,
              Gower Street,
              London WC1E 6BT, UK
         \and
              School of Physics \& Astronomy,
              E.C. Stoner Building,
              The University of Leeds,
              Leeds, LS2 9JT, UK
             }

   \date{Received <date>; accepted <date>}

% \abstract{}{}{}{}{} 
% 5 {} token are mandatory
 
  \abstract
   {Massive binaries have stellar winds that collide. In the colliding-wind
    region, various physically interesting processes occur, leading to
    enhanced X-ray emission, non-thermal radio emission, as well as 
    non-thermal X-rays and gamma-rays. Non-thermal radio emission 
    (due to synchrotron radiation) has
    so far been observed at centimetre wavelengths.
    At millimetre wavelengths, the stellar winds and the colliding-wind
    region emit more thermal free-free radiation,
    and it is expected that
    any non-thermal contribution will be difficult or impossible to detect.
   }
   {We aim to determine if the material in the colliding-wind region
    contributes substantially to the observed millimetre fluxes of a 
    colliding-wind binary. We also try to distinguish the synchrotron emission
    from the free-free emission.
   }
   {We monitored the massive binary \object{Cyg~OB2~\#8A} at 3~mm with the
    NOrthern Extended Millimeter Array (NOEMA) interferometer of the 
    Institut de Radioastronomie Millim\'etrique (IRAM). The data were
    collected in 14 separate observing runs (in 2014 and 2016), 
    and provide good coverage of the orbital period.
   }
   {The observed millimetre fluxes range between 1.1 and 2.3~mJy,
    and show phase-locked variability, clearly
    indicating that a large part of the emission is due to the 
    colliding-wind region.
    A simple synchrotron model gives 
    fluxes with the correct
    order of magnitude, but with a maximum that is phase-shifted
    with respect to the observations.
    Qualitatively this
    phase shift can be explained by our neglect of orbital motion
    on the shape of the colliding-wind region.
    A model using only free-free emission
    results in only a slightly worse explanation of the observations.
    Additionally, on the map of our observations we also detect the 
    O6.5 III star \object{Cyg~OB2~\#8B}, for which we
    determine a 3~mm flux of 0.21 $\pm$ 0.033~mJy.
   }
   {The question of whether synchrotron radiation or free-free emission
    dominates the millimetre fluxes of \object{Cyg~OB2~\#8A} remains open.
    More detailed modelling of this system, 
    based on solving the hydrodynamical
    equations, is required to give a definite answer.}

   \keywords{Binaries: spectroscopic -
             Stars: winds, outflows -
             Stars: individual: Cyg~OB2~\#8A -
             Stars: individual: Cyg~OB2~\#8B -
             Stars: massive -
             Radio continuum: stars}
   \maketitle
%
%-------------------------------------------------------------------

\section{Introduction}

Massive stars have strong stellar winds. When two such stars form a binary,
their stellar winds collide, leading to a number of interesting effects.
The collision between the two winds heats up the material to such temperatures 
that it emits detectable X-ray radiation,
as originally predicted by
\citet{1976SvA....20....2P} and recently reviewed by
\citet{2016AdSpR..58..761R}. 
In the shocks associated with the colliding-wind region (CWR),
a fraction of the electrons are accelerated to high, non-thermal energies by the Fermi mechanism
\citep{1978MNRAS.182..147B, 1993ApJ...402..271E}.
These electrons spiral in the magnetic field, thereby emitting synchrotron
emission, which we can detect as non-thermal radio emission 
\citep{1989ApJ...340..518B}.

In an eccentric binary, the strength of the collision (more specifically, the ram
pressure) will vary with orbital phase, leading to phase-locked variations of the intrinsic emission 
coming from the CWR. At radio wavelengths, there is furthermore the effect
of the strong free-free absorption in the stellar wind material \citep{1975MNRAS.170...41W}.
The changing positions of both stellar winds and the CWR lead to
further phase-locked variations in the observed radio fluxes.
This has been observed in a number of O+O and Wolf-Rayet+O
colliding-wind binaries
\citep[e.g.,][]{1995ApJ...451..352W,
2005A&A...436.1033B,
2007A&A...464..701B,
2014A&A...561A..18B}.
In exceptional cases, the CWR has been resolved by
high-resolution radio observations
\citep[][and references therein]{2015A&A...579A..99B}.

\begin{table*}
\caption{Observing log of the NOEMA 3~mm observations of \object{Cyg~OB2~\#8A}.}
\label{table observing log}
\centering
\begin{tabular}{rrcl}
\hline\hline
\multicolumn{2}{c}{Observation} & \multicolumn{1}{c}{Number} & \multicolumn{1}{c}{Antenna} \\
\multicolumn{1}{c}{Date} & \multicolumn{1}{c}{Time (UT)} & \multicolumn{1}{c}{of obs.\tablefootmark{a}} & \multicolumn{1}{c}{Configuration} \\
\hline
2014-11-24 & 18:50--20:30 & 3 & W09E04W12N17N11E10 \\
2014-11-30 & 20:47--23:21 & 4 & W09E04W12N17N11E10 \\
2014-12-01 & 18:21--21:51 & 6 & W09E04W12N17N11 \\[1.0ex]
2016-06-01 & 06:02--06:27 & 1 & N02W12E04N11E10N07 \\
2016-06-03 & 02:02--03:35 & 2 & N02W12W09E04N11E10N07 \\
2016-06-06 & 04:16--06:16 & 2 & W09N17E04N29E10N20 \\
2016-06-08 & 04:12--06:04 & 2 & W09N17E04N29E10N20 \\
2016-06-10 & 04:01--05:18 & 2 & N17E04N29E10N20 \\
2016-06-15 & 07:16--08:38 & 2 & W12W09N17E04N29E10N20 \\
2016-06-17 & 20:13--20:39 & 2 & W09N17E04N29E10N20 \\
2016-06-20 & 04:14--07:15 & 2 & W09N17N29E10N20 \\
2016-06-22 & 20:38--21:50 & 2 & E04W12W09N17N29E10N20 \\
2016-06-25 & 01:06--03:08 & 2 & E04W12W09N17N29E10N20 \\
2016-07-07 & 23:12--24:29 & 2 & E04W09N17N29E10N20 \\
\hline
\end{tabular}
\tablefoot{
\tablefoottext{a}{A single observation on Cyg~OB2~\#8A consists of 30 or 33 scans
(for the 2014 and 2016 runs, respectively), where each scan is 45 s.}
}
\end{table*}

Colliding
winds also emit non-thermal X-rays and gamma-rays
\citep{1987A&A...171..135P,
2003A&A...399.1121B,
2006A&A...446.1001P,
2006ApJ...644.1118R,
2007A&ARv..14..171D,
2014ApJ...789...87R},
making these objects very relevant to high-energy
physics \citep{2006MNRAS.372..801P}. 
The regime of physical parameters (density, magnetic field, ambient
radiation field, etc.) is quite different from that of other high-energy environments, and colliding-wind
binaries can therefore provide important tests of our understanding of the physical processes responsible.

Colliding-wind systems can also be highly relevant for the mass-loss rate determinations in single stars.
This has become a major problem in massive star research in recent years, due to the fact that winds
are clumped and porous. Taking into account clumping decreases the mass loss rates by a factor 3-10
compared to what was previously thought
\citep[e.g.,][]{2008A&ARv..16..209P}.
This, in turn, has major consequences
for stellar and galactic evolution. Colliding-wind systems can provide an independent determination of
the effect of clumping and porosity
\citep[e.g.][]{2007ApJ...660L.141P}: 
by modelling the colliding winds, and predicting the 
various observational indicators, we can constrain the mass-loss rate and/or porosity
in the wind. 

Colliding-wind binaries have so far been relatively unexplored at millimetre wavelengths. 
This is mainly
due to the expectation that free-free emission will start to dominate
the fluxes at shorter wavelength, and that therefore the non-thermal 
synchrotron emission will be difficult or impossible to detect.

Furthermore, the CWR will also increase the free-free emission,
as shown by 
\citet{1995MNRAS.277..163S} and \citet{2011A&A...531A..52M}, 
using simplified models for a radiative shock;
and by 
\citet{2003A&A...409..217D},
\citet{2006A&A...446.1001P} and
\citet{2010MNRAS.403.1633P}, 
using hydrodynamical calculations for both adiabatic
and radiative shocks.
The density increase leads to a considerable
amount of additional free-free emission.
\citet{2015RMxAA..51..209M} tried to look for this,
by comparing centimetre and millimetre fluxes
for a sample of 17 Wolf-Rayet stars. For three of these, a hint from the colliding-wind
region contribution was found, but their observations did not allow to clearly identify
the physical mechanism.

Another approach is to look for the variability of the millimetre fluxes.
An eccentric binary will show
phase-locked variability of the millimetre fluxes for the same reason
as for the radio fluxes. The changing distance between the two components
leads to variation in the strength of the collision and hence the amount of
additional free-free emission. Added to that is the variation in the free-free 
absorption due to the changing position
of the stellar winds and the CWR.

\object{Cyg~OB2~\#8A} (= Schulte 8A = MT 465 = BD+40 4227A)
is a well-known massive colliding-wind binary, and
therefore an appropriate 
candidate to look for variability at millimetre wavelengths.
It is a member of \object{Cyg~OB2},
which is an association containing a large number of massive
stars \citep{2000A&A...360..539K,2015MNRAS.449..741W}. 
The association 
harbours a number of massive colliding-wind binaries that have
been studied in detail
\citep{2008A&A...483..585V,
2010A&A...519A.111B,
2010ApJ...709..632K,
2013A&A...550A..90B}.

\object{Cyg~OB2~\#8A} was discovered to be a binary
by \citet{2004A&A...424L..39D}.
It has a period $P$ = 21.908 $\pm$ 0.040 d, an eccentricity 
$e$ = 0.24 $\pm$ 0.04,
and it consists of an O6If primary and an O5.5III(f) secondary
\citep{2006MNRAS.371.1280D}. It has all the attributes of a colliding-wind
binary. Its radio spectrum shows a negative spectral index
\citep{1989ApJ...340..518B} and the radio flux variations are locked
to the orbital phase \citep{2010A&A...519A.111B}. The X-ray emission
is overluminous compared to that of single O-type stars and its
variability is also phase-locked
\citep{2006MNRAS.371.1280D,2010A&A...519A.111B}.
Among the O-type colliding-wind binaries, it is the system where 
the parameters of the stars, their stellar wind and their orbit are best known
\citep{2007A&ARv..14..171D}. 

We observed \object{Cyg~OB2~\#8A} with the NOrthern Extended Millimeter Array (NOEMA)
interferometer of the Institut de Radioastronomie Millim\'etrique 
(IRAM\footnote{\url{http://www.iram-institute.org/}}),
to see if we could detect the expected flux variations
at millimetre wavelengths.
We present the observations in
Sect.~\ref{section observations} and the data reduction in
Sect.~\ref{section data reduction}. 
The resulting millimetre
light curve is given in Sect.~\ref{section millimetre light curve}
and we model it in 
Sect.~\ref{section modelling}. 
In Sect.~\ref{section 8B}, we present the results for 
\object{Cyg~OB2~\#8B}, which is the visual companion to our main target.
The conclusions are given
in Sect.~\ref{section conclusions}.

\section{Observations}
\label{section observations}

\begin{table*}[t]
\caption{Measured 3~mm fluxes of Cyg~OB2~\#8A.}
\label{table results}
\centering
\begin{tabular}{rrrrrccr@{ $\times$ }rrr}
\hline\hline
\multicolumn{1}{c}{(1)} & \multicolumn{1}{c}{(2)} &\multicolumn{1}{c}{(3)} &\multicolumn{1}{c}{(4)} &\multicolumn{1}{c}{(5)} &\multicolumn{1}{c}{(6)} &\multicolumn{1}{c}{(7)} &\multicolumn{1}{c}{(8)} &\multicolumn{1}{c}{(9)} &\multicolumn{1}{c}{(10)} &\multicolumn{1}{c}{(11)}\\
\multicolumn{1}{c}{Observation}  & \multicolumn{1}{c}{JD} & \multicolumn{1}{c}{Orbital} & \multicolumn{2}{c}{Quality (\%)} & \multicolumn{1}{c}{Flux (Jy)} & \multicolumn{1}{c}{Flux (Jy)} & \multicolumn{3}{c}{Beam} & \multicolumn{1}{c}{Flux} \\\cline{4-5}\cline{8-10}
\multicolumn{1}{c}{Date} & \multicolumn{1}{c}{$-$2\,450\,000.0} & \multicolumn{1}{c}{Phase} & \multicolumn{1}{c}{Ampl} & \multicolumn{1}{c}{Phase}                 & \multicolumn{1}{c}{2200+420} & \multicolumn{1}{c}{2005+403} & \multicolumn{1}{c}{major} & \multicolumn{1}{c}{minor} & \multicolumn{1}{c}{PA} & \multicolumn{1}{c}{(mJy)} \\
\hline
2014-11-24 & 6986.3194 & 0.41 &   0 &   0 & 4.39 & 0.52 &  2.82\arcsec & 2.30\arcsec &   45\fdg5 & 1.28 $\pm$ 0.17 \\
2014-11-30 & 6992.4194 & 0.68 &  11 &  12 & 5.01 & 0.49 & 11.68\arcsec & 3.26\arcsec &   19\fdg5 & 2.30 $\pm$ 0.42 \\
2014-12-01 & 6993.3375 & 0.73 &   0 &   3 & 4.90 & 0.57 &  3.61\arcsec & 2.21\arcsec &   67\fdg5 & 2.24 $\pm$ 0.19 \\
           &           &      &     &     & \multicolumn{2}{c}{2013+370} \\
2016-06-01 & 7540.7601 & 0.71 &   0 &   3 & \multicolumn{2}{c}{3.50} &  4.89\arcsec & 2.87\arcsec &$-$158\fdg0 & 1.64 $\pm$ 0.26 \\
2016-06-03 & 7542.6170 & 0.80 &   0 &   0 & \multicolumn{2}{c}{3.61} &  4.31\arcsec & 2.42\arcsec &    23\fdg3 & 2.22 $\pm$ 0.10 \\
2016-06-06 & 7545.7194 & 0.94 &   0 &   0 & \multicolumn{2}{c}{3.68} &  2.99\arcsec & 1.34\arcsec &    59\fdg7 & 1.99 $\pm$ 0.11 \\
2016-06-08 & 7547.7139 & 0.03 &   0 &   5 & \multicolumn{2}{c}{3.68} &  2.99\arcsec & 1.48\arcsec &    59\fdg7 & 1.65 $\pm$ 0.13 \\
2016-06-10 & 7549.6941 & 0.12 &   0 &   0 & \multicolumn{2}{c}{3.43} & 15.79\arcsec & 9.29\arcsec & $-$43\fdg8 & 1.27 $\pm$ 0.11 \\
2016-06-15 & 7554.8313 & 0.36 &  13 &   6 & \multicolumn{2}{c}{2.31} & \multicolumn{4}{c}{No detection} \\
2016-06-17 & 7557.3514 & 0.47 &   0 &  50 & \multicolumn{2}{c}{3.15} & \multicolumn{4}{c}{No detection} \\
2016-06-20 & 7559.7392 & 0.58 &   0 &  15 & \multicolumn{2}{c}{3.53} &  8.83\arcsec & 2.84\arcsec &    80\fdg7 & 1.67 $\pm$ 0.15 \\
2016-06-22 & 7562.3847 & 0.70 &  13 &  43 & \multicolumn{2}{c}{3.32} & 17.24\arcsec & 4.71\arcsec &    79\fdg3 & 1.10 $\pm$ 0.31 \\
2016-06-25 & 7564.5882 & 0.80 &   0 &  12 & \multicolumn{2}{c}{3.31} &  2.20\arcsec & 1.66\arcsec &    83\fdg9 & 1.89 $\pm$ 0.10 \\
2016-07-07 & 7577.4934 & 0.39 &   0 &   7 & \multicolumn{2}{c}{3.29} &  3.23\arcsec & 1.41\arcsec &    93\fdg2 & 1.35 $\pm$ 0.13 \\
\hline
\end{tabular}
\tablefoot{
Column (1) gives the date of the observation;
column (2) the Julian date of the mid-point of the observation;
column (3) the phase in the 21.908 d orbital period;
column (4) and (5) the percentage of data that were flagged because of 
high amplitude loss and high phase loss, respectively;
column (6) and (7) the fluxes of the phase calibrators (only one calibrator for the 2016 data);
column (8), (9) and (10) the synthesized beam (major axis $\times$ minor axis and position angle PA);
column (11) the flux of Cyg~OB2~\#8A, and its error bar.
}
\end{table*}

\object{Cyg~OB2~\#8A} was observed with the IRAM NOEMA interferometer
during two different periods: 
end of 2014 (project: S14AW; PI: D.M. Fenech) and 
mid-2016 (project S16AU; PI: R. Blomme).
The 2014 observations consist of three runs (a fourth run, on 2014-11-29,
contains only calibrator data and no
on-target observations, so it will not be considered further).
The 2016 observations have 11 runs.
The detailed observing log is given in Table~\ref{table observing log}.

Observations were collected at 3~mm (110 GHz; IRAM receiver no.~1 in
Upper Side Band),
in single-pointing mode. Dual polarization was used with a
bandwidth of $2 \times 3.6$ GHz.
An observing run consists of alternating between one or two phase 
calibrators and \object{Cyg~OB2~\#8A}, ensuring that each \object{Cyg~OB2~\#8A} observation
is preceded and succeeded by a phase calibrator observation. In addition,
for each run, the flux calibrator \object{MWC~349} and the bandpass
calibrator \object{3C454.3} were observed (for the 2016-06-22 observation
the bandpass calibrator 3C273 was also observed).

During the 2014 runs, the two phase calibrators used were
\object{2200+420} (at an angular distance $\Delta=$ 16{\fdg}7)
and \object{2005+403} ($\Delta=$ 4{\fdg}9);
in 2016, only \object{2013+370}  ($\Delta=$ 5{\fdg}4) was used.
A single \object{Cyg~OB2~\#8A} observation consists of 30 scans 
(33 scans in 2016)
of 45 s each. For each phase calibrator observation, three 45 s scans were done.
The signals from the antennas were correlated with the WideX correlator.
During the 2014 observations, the antennas were in the 6Cq configuration
(with E10 missing for the 2014-12-01 run). For the 2016 observations,
many changes in antenna positions were made; we therefore list the detailed
configurations in Table~\ref{table observing log}.

\section{Data reduction}
\label{section data reduction}

We used the 
GILDAS\footnote{\url{http://www.iram.fr/IRAMFR/GILDAS/}} software for
the data reduction, specifically the packages CLIC (for calibration)
and MAPPING (for the flux determination).

We split the observing runs, so that we can separately calibrate 
each series of 30 (or 33 scans) on \object{Cyg~OB2~\#8A}. The phase calibrator(s)
immediately preceding and following the \object{Cyg~OB2~\#8A} observations
were selected, as well as the bandpass and flux calibrators that are
nearest in time. Based on the pipeline reduction that had run at IRAM,
we flagged some data which were clearly discrepant, or were on one
side of a cable phase jump, or had high system temperatures.
The 2014 observations were done in a stable configuration, and therefore
needed no baseline corrections. Antenna
movement was frequent however during the 2016 observations, and for all
these runs, baseline corrections (provided by IRAM staff) were applied.

Default automatic flagging was applied, which flags problems with
antenna shadowing, data that do not have
surrounding calibrator observations, and timing errors.
Next, atmospheric phase corrections were applied. 
The receiver bandpass was then calibrated
on the bandpass calibrator \object{3C454.3}, using the
default degree of the polynomials. For the 2016-06-22 observations,
the \object{3C454.3} data were not usable, and the bandpass was calibrated
on \object{3C273}. Phase calibration consists
of fitting a per-antenna linear function to the phases of the 
phase calibrator(s). Fluxes were then calibrated on the flux
calibrator \object{MWC 349}, which has a flux of 1.23 Jy. For the
amplitude calibration, a per-antenna linear function was fitted
to the calibrators. 

As the final part of the calibration, the
Data Quality Assessment procedure was run. This flags data with
more than 40\degr~root-mean-square (RMS) of phase loss, and more than 20 \%
of amplitude loss. The fraction of points that were flagged 
in this way are indicated in Table~\ref{table results}. They
are a good indicator of the quality of the data, in the sense
that a higher fraction of flagged points indicates worse quality.
Pointing, focus and tracking errors which
are too high are also flagged, but this did not occur for any
of our observations. Finally, the calibrated data of \object{Cyg~OB2~\#8A} were 
written out. At this stage, all \object{Cyg~OB2~\#8A} data taken during
a single observing run were merged again.

\begin{figure}
\resizebox{\hsize}{!}{\includegraphics[bb=15 45 630 545,clip]{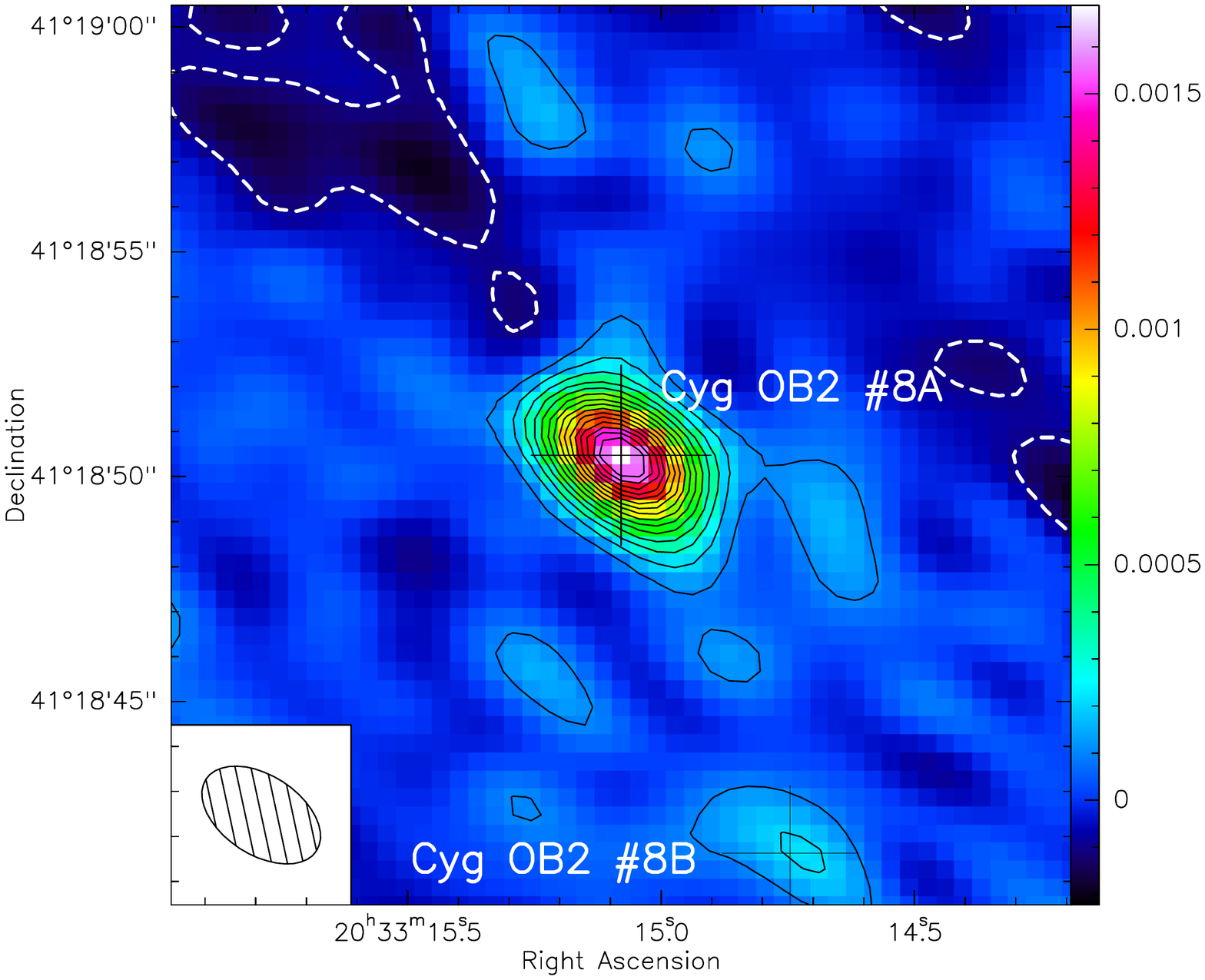}}
\caption{Deep 3~mm image of all \object{Cyg~OB2~\#8A} data combined. 
The colour scale values are in Jy. The contour levels
are at $-0.1$~mJy, $+0.1$~mJy and then go up in steps of 
0.1~mJy to 1.6~mJy. The negative contour is indicated by the dashed, white line.
The crosses indicate the positions
of \object{Cyg~OB2~\#8A} and \#8B (from SIMBAD).
The synthesized beam (shown in the lower left corner) is
$2.93\arcsec \times 1.76\arcsec$ with a position angle of 57\fdg1.
}
\label{fig deep image}
\end{figure}

Next, we made a deep image by combining all the data
(Fig.~\ref{fig deep image}), to see if there are any
other sources in the field besides \object{Cyg~OB2~\#8A}.
We used a pixel size of 0.4\arcsec $\times$ 0.4\arcsec,
which is about one-quarter (linear dimension) of the synthesized beam size. We took
a grid size of 256 $\times$ 256 pixels, which covers somewhat more than 
twice the primary beam and we applied natural weighting. 
We then cleaned the image, 
stopping when we reached 0.025~mJy, which is half of the RMS noise level
in the image. Fig.~\ref{fig deep image} shows only the inner 20\arcsec $\times$ 
20\arcsec~part of this image. 

The field is dominated by \object{Cyg~OB2~\#8A}, which is -- as expected -- 
a point source. A much weaker source is seen at the bottom edge of 
Fig.~\ref{fig deep image}. We identify it as \object{Cyg~OB2~\#8B},
and discuss it further in Sect.~\ref{section 8B}.
We checked the full image for other possible sources, but found none.

As the data are dominated by the flux from \object{Cyg~OB2~\#8A},
we determined
its flux by fitting a point source model to the observed visibilities
for each observing run.
This is a better approach than measuring the flux on images made
from these visibilities, as the cleaning procedure 
might introduce artefacts.
The fluxes we obtained are listed in Table~\ref{table results}.

A number of variant reductions were also tried: switching off the
atmospheric phase correction, using weights
in the phase or amplitude calibration, or averaging the polarizations
in these calibrations, or flagging frequency parasites
(i.e. sharp peaks in frequency caused by the instrument).
These all gave values within the listed error bars. 
As the
\object{Cyg~OB2~\#8A} flux is relatively high, we also tried to include it
in the calibration process, but this did not improve results.

\begin{table*}
\caption{Overview of models used in Sect.~\ref{section modelling}.}
\label{table models}
\centering
\begin{tabular}{cl}
\hline\hline
Section & Model \\
\hline
\ref{section stellar wind contribution} &
Spherically symmetric wind; thermal wind emission only. \\
\ref{section CWR synchrotron emission} &
Detailed synchrotron emission model; no increased temperature of CWR
material; thermal wind emission. \\
\ref{section adiabatic model} &
Increased temperature of CWR material; adiabatic shock; no synchrotron emission; thermal wind emission. \\
\ref{section radiative model} &
Increased temperature of CWR material; radiative shock; no synchrotron emission; thermal wind emission. \\
\ref{section combined model} &
Detailed synchrotron emission model; increased temperature of CWR
material; thermal wind emission. \\
\hline
\end{tabular}
\end{table*}

\begin{table}
\caption{Cyg~OB2~\#8A parameters used in modelling.}
\label{table parameters}
\centering
\begin{tabular}{llllll}
\hline\hline
Parameter & Primary & Secondary \\
\hline
$T_{\rm eff}$ (K)                & 36800 & 39200 \\
log $L_{\rm bol}/L_{\sun}$         & 5.82  & 5.67 \\
$\dot{M}$ ($M_{\sun} {\rm yr}^{-1}$) & $4.8 \times 10^{-6}$ & $3.0 \times 10^{-6}$ \\
$\varv_\infty$ (km s$^{-1}$)           & 1873  & 2107  \\
3~mm flux (mJy) &  1.62 & 0.75 \\
3~mm $\tau=1$ radius ($R_{\sun}$)    & 220 & 140 \\
semi-major axis $a$ ($R_{\sun}$) & 65.7 & 76.0 \\
period $P$ (days)  & \multicolumn{2}{c}{$21.908 \pm 0.040$} \\
eccentricity $e$   & \multicolumn{2}{c}{$0.24 \pm 0.04$} \\
inclination $i$    & \multicolumn{2}{c}{$32\degr \pm 5\degr$} \\
distance $D$ (kpc) & \multicolumn{2}{c}{1.4} \\
\hline
\end{tabular}
\tablefoot{The parameters were derived by \citet{2006MNRAS.371.1280D}.
The predicted fluxes and the radii where the optical depth $\tau$
becomes 1 for 3~mm ($R_{\tau=1}$) have been derived from the \citet{1975MNRAS.170...41W}
equations (see Sect.~\ref{section stellar wind contribution})
The values listed do not include clumping. When including clumping,
the flux scales as $(\dot{M} \sqrt{f_{\rm cl}})^{4/3}$ and 
$R_{\tau=1}$
scales differently as $(\dot{M} \sqrt{f_{\rm cl}})^{2/3}$,
where $f_{\rm cl}$ is the clumping factor. Note, however, that when comparing
a clumped wind and a smooth wind that have the same flux, their 
$\dot{M} \sqrt{f_{\rm cl}}$ needs to be the same, hence also their $R_{\tau=1}$.
}
\end{table}

\section{Millimetre light curve}
\label{section millimetre light curve}

The \object{Cyg~OB2~\#8A} fluxes listed in Table~\ref{table results} clearly
indicate variability. To check that this is not some artefact of the
data reduction, we also list the fluxes of the phase calibrators.
The phase calibrators are intrinsically variable, and their measured
fluxes are clearly not correlated with the \object{Cyg~OB2~\#8A} ones.
We can therefore exclude data reduction artefacts.
The absolute flux calibration at 3~mm is good to about 10 \%
(J. M. Winters, pers. comm.), which corresponds to about $\pm$ 0.2~mJy. 
The flux variations are therefore also not due to 
changes in the levels of the absolute flux calibration.

\begin{figure}
\resizebox{\hsize}{!}{\includegraphics[bb=10 20 400 300,clip]{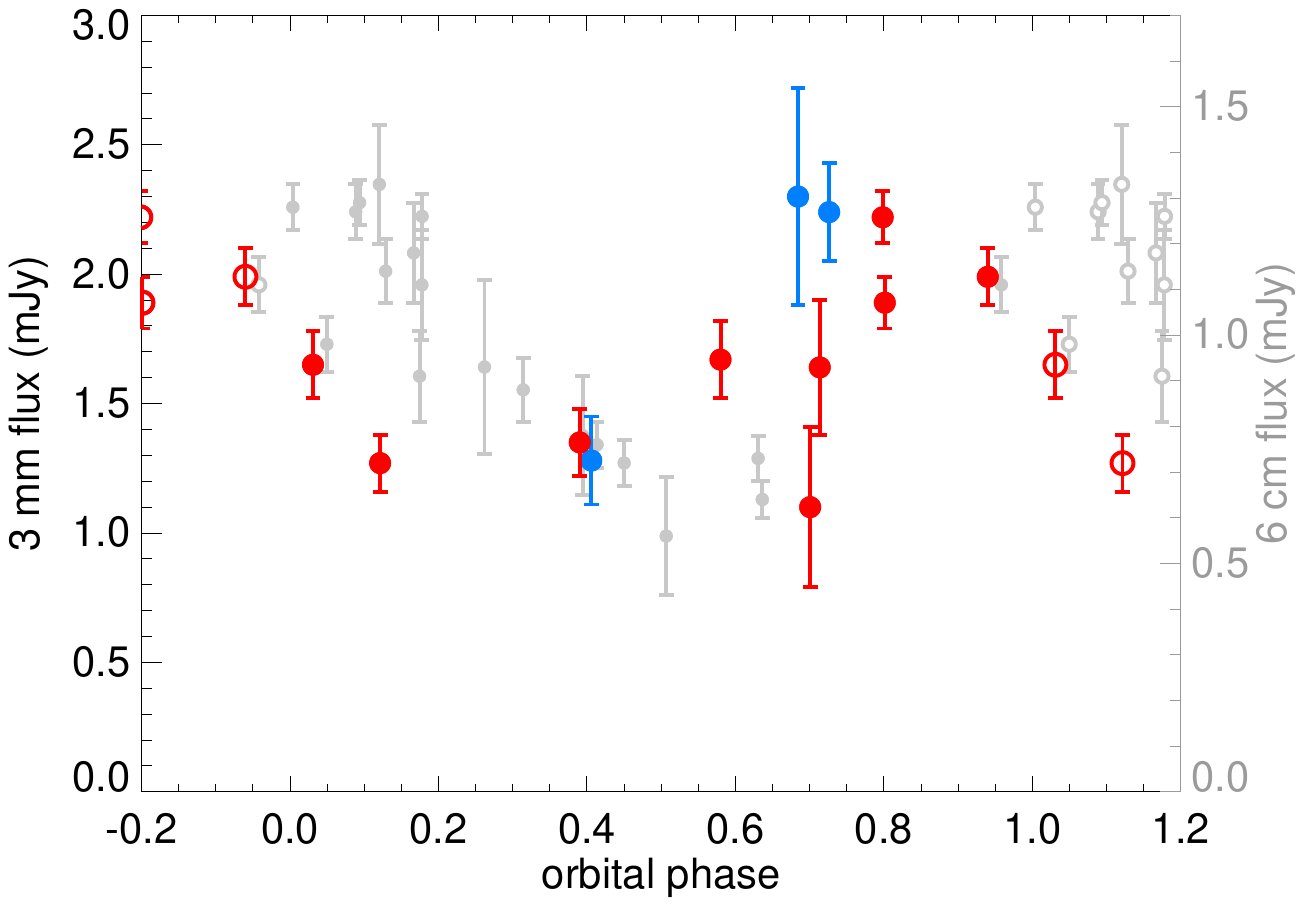}}
\caption{Observed 3~mm fluxes of \object{Cyg~OB2~\#8A}, plotted as a function of orbital
phase in the 21.908-day binary period. The blue symbols show the 2014 data,
the red ones the 2016 data. To better show the behaviour of the light
curve, the phase range is 
extended by 0.2 on each side. Open symbols indicate duplications in that extended
range. The grey data show the 6~cm observations from \citet{2010A&A...519A.111B}.
Note that the 3~mm flux scale (left) is different from the 6~cm one (right).}
\label{fig fluxes}
\end{figure}

We next plot the observed fluxes in the orbital phase diagram
(Fig.~\ref{fig fluxes}). The fluxes are clearly correlated with
the orbital phase, indicating that they are at least partly
formed in the CWR.
Around phase 0.7, there is a large range in the fluxes,
which seems to violate the assumption that, in a colliding-wind binary,
the fluxes repeat nearly perfectly from one orbit to another.
We note, however, that the two extreme fluxes are of lower quality:
the amplitude quality indicator in Table~\ref{table results} shows that
11 and 13 \% of the data for these two observations have been
removed by the Data Quality
Assessment procedure, because they show more than 20 \% amplitude loss.
Note that the standard error bar does not include this systematic effect.
On the basis of this quality information, we consider the flux variations
around phase ~0.7 to be not significant.

Fig.~\ref{fig fluxes} also shows the 6~cm radio fluxes from
\citet{2010A&A...519A.111B}. Note that different flux scales are
used for the 3~mm and the 6~cm data. The shape of the
3~mm light curve follows that of the 6~cm one very well, 
but differs from it in two aspects.
First, the millimetre data are above the median (1.66 mJy) about 30 \%
of the time, while the 6~cm data are above their median about 50 \%
of the time. This is due to the non-uniform distribution of
the phases for which we have flux determinations. There are a large
number of observations around the phase of maximum flux (0.7 -- 0.8),
but none around the suspected minimum (phase 0.2 -- 0.3). Additional
data around the minimum would lower the median and lead to a
more equal distribution of the phase range above and below the median.

Secondly, the 3~mm light curve is phase-shifted with respect
to the 6~cm one.
One could consider that the uncertainty in
the period is responsible for this shift. The $21.908 \pm 0.040$~d
period was derived by \citet{2004A&A...424L..39D} on the basis
of spectroscopic data from the years 2000 -- 2003. However,
both X-ray data covering the years 1991 -- 2004 and radio data
covering the years 1980 -- 2005 are consistent with this
period, suggesting that the $0.040$~d error bar is too conservative
\citep{2006MNRAS.371.1280D, 2010A&A...519A.111B}.
We therefore consider it unlikely that the phase shift
is due to uncertainty in the period.

Alternatively, we note that a phase shift in the 
radio light curves at different wavelengths is quite common
for colliding-wind binaries. The most detailed example is
\object{WR 140}, as shown by \citet{1995ApJ...451..352W}.
Specifically for \object{Cyg~OB2~\#8A}, a hint of such a phase
shift is present in Fig.~1 of 
\citet{2010A&A...519A.111B}. It can be seen when comparing the position of minimum
for the 6~cm and 3.6~cm radio observations.

\section{Modelling}
\label{section modelling}

In the following sections a number of different models will be applied
to the observed data. Table~\ref{table models} summarizes these
models, allowing the reader to keep track of which model is used in which 
section.

\subsection{Stellar wind contribution}
\label{section stellar wind contribution}

We first estimate how much the two stellar winds contribute to the
observed millimetre fluxes. As a first approximation, we apply the
\citet{1975MNRAS.170...41W} formalism. The expected flux 
$S_\nu$ (in mJy) at frequency $\nu$ (in Hz) of a
spherically symmetric, steady-state wind, with a mass-loss rate 
$\dot{M}$ (in ${\rm M}_{\sun}{\rm yr}^{-1}$) flowing out
at a constant velocity 
$\varv_\infty$ (in ${\rm km\,s}^{-1}$) is:
\begin{equation}
S_\nu = 23.2\times 10^3 \left( \frac{\dot{M}}{\mu \varv_\infty} \right)^{4/3}
        \frac{\nu^{2/3}}{D^2} \left(\gamma g_{\rm ff}(\nu,T_{\rm e}) \overline{Z^2}\right)^{2/3},
\label{equation Wright Barlow}
\end{equation}
where $\mu$ is the mean atomic weight (1.27 for solar
composition), $D$ is the distance (in kpc),
$\overline{Z^2}$ is the mean squared ion charge and the Gaunt factor
$g_{\rm ff}$ at frequency $\nu$ and electron temperature
$T_{\rm e}$ is given by \citep{1991ApJ...377..629L}:
\begin{equation}
g_{\rm ff}(\nu,T_{\rm e}) = 9.77 \left(1+0.13 \log_{10} \frac{T_{\rm e}^{3/2}}{\sqrt{\overline{Z^2}} \nu} \right)
\end{equation}

We apply these equations using the stellar wind parameters 
which were derived by \citet{2006MNRAS.371.1280D}, and which are also
listed in Table~\ref{table parameters}.
For $T_{\rm e}$, we take half the effective temperature of the star.
The distance to the \object{Cyg~OB2} association is not well known. We take
$D=1.4$~kpc, following \citet{2016MNRAS.463..763M}, who give a detailed
overview of the various distance determinations. Using these values,
we find a 3~mm flux of 1.62~mJy for the primary, and 0.75~mJy for
the secondary.

A basic assumption of the \citet{1975MNRAS.170...41W} model is that
the wind is flowing out at a constant velocity. Because of the
wavelength-squared dependence of the free-free absorption,
the millimetre fluxes are formed closer to the star than the
centimetre fluxes. In this formation region, the velocity will
not necessarily have reached its terminal value,
and this will lead to a different value for the flux
\citep[see, e.g.,][]{2010MNRAS.403.1633P,2016MNRAS.463.2735D}. We therefore
also made a numerical integration of the specific intensities and
the flux in a spherically
symmetric wind, assuming a $\beta=0.8$ velocity law. The effect
for our specific set of stellar wind parameters is minor
however: we now obtain 1.63~mJy for the primary, and 0.76~mJy
for the secondary. Larger differences occur at shorter wavelengths
where the wind is probed still deeper
\citep{2010MNRAS.403.1633P}.

The summed flux of both stars gives 2.39~mJy, which is comparable
to the highest flux value observed. This may seem surprising,
but it should be realized that we are no longer dealing with
spherically symmetric winds. In the overlap region between the two
winds, the wind of the other component is ``missing" in the
sense that the material
has been compressed into the CWR. 
Models presented in the following sections will take this effect into
account.

For use in the following sections, we also derive the radius
at which the optical depth of 1 is reached. Using the same 
assumptions as \citet{1975MNRAS.170...41W}, it can be shown that:
\begin{equation}
R_{\tau=1} = \left(\frac{K(\nu,T)}{3}\right)^{1/3} \left(\frac{\dot{M}}{4\pi \mu m_{\rm H} \varv_\infty}\right)^{2/3},
\end{equation}
where
\begin{equation}
K(\nu,T_{\rm e}) = 3.7\times 10^8 \left\{1-\exp\left(-\frac{h \nu}{k T_{\rm e}}\right)\right\} \frac{Z^2 g_{\rm ff}(\nu,T_{\rm e})}{T_{\rm e}^{1/2} \nu^3}.
\end{equation}
Filling in the numbers we find $R_{\tau=1} = 220~R_\sun$ and $140~R_\sun$
for the primary and secondary, respectively.

\subsection{CWR synchrotron emission}
\label{section CWR synchrotron emission}

We first estimate if synchrotron emission can still be a contributing
factor at 3~mm. A relativistic electron with energy $E$
will emit synchrotron radiation over a range of frequencies.
The frequency $\nu_{\rm max}$
(in MHz) at which the
maximum emission occurs is given by 
\citep[][their Eq.~2.23]{1965ARA&A...3..297G}:
\begin{equation}
\nu_{\rm max} = 1.2 B \sin \theta \left(\frac{E}{m_{\rm e}c^2}\right)^2,
\end{equation}
where $B$ is the local magnetic field (in G),
$\theta$ is the angle between the velocity vector of the electron
and the magnetic field vector,
$m_{\rm e}$ is the mass of the electron,
and $c$ is the speed of light.
This equation neglects the 
Razin effect \citep[][their Sect.~4]{1965ARA&A...3..297G},
which decreases the flux at longer wavelengths.

Taking an order-of-magnitude value of 0.1 G for $B$, and neglecting
the $\sin \theta$ factor, we find that 3~mm (110~GHz) emission
is mainly due to electrons with an energy of $E \approx 450$~MeV.
To know if such electrons exist in the CWR, we need to balance the
Fermi acceleration against inverse Compton cooling,
which is the main cooling mechanism. Inverse Compton cooling is
due to the interaction of the relativistic electrons with the
high-energy photons from both stars.
\citet{2005A&A...433..313V} derive an expression for the highest 
energy that electrons can attain under these conditions:
\begin{equation}
\left(\frac{E}{m_{\rm e} c^2}\right)^2 = 
   \frac{\chi_{\rm s} \Delta u^2}{\chi_{\rm s}^2-1}
   \frac{4 \pi r^2}{\sigma_{\rm T} L_*}
   \frac{eB}{c},
\end{equation}
where 
$e$ is the charge of the electron,
$\chi_{\rm s}$ is the shock strength, $\Delta u$ is the velocity jump,
$\sigma_{\rm T}$ the Thomson electron scattering cross section,
and $L_*$ the luminosity of the star.
For an estimate, we take values which are roughly applicable to 
\object{Cyg~OB2~\#8A} at periastron: $r=100~R_\sun$, $\log L_*/L_\sun = 5.7$,
$B$ = 0.1 G,
and for the shock: $\chi_{\rm s}=4$ and $\Delta u = 1500~{\rm km\,s}^{-1}$). 
This
gives $E \approx 1000$~MeV. There are therefore electrons in the CWR
that have sufficient energy to emit synchrotron radiation at 3~mm.

To investigate the synchrotron emission in more detail, we used
the model from
\citet{2010A&A...519A.111B}. This model follows in detail the momentum 
distribution of the relativistic electrons as they move away from the shock, 
taking into account inverse Compton and adiabatic cooling.
A simplifying assumption of the model is that
it does not solve the hydrodynamical equations, but it
uses an analytic approach to determine the position of the shocks.
From the momentum distribution, the synchrotron emissivity is
calculated. This is then mapped
into a three-dimensional grid (assuming rotational symmetry around
the line connecting the two stars). In this grid, the radiative transfer
equation is solved, taking into account the synchrotron emission
as well as free-free absorption and emission due to the stellar winds.
No free-free contribution from the CWR is included.
The model is applied at a number of orbital phases, and the
theoretical flux is determined for each of these phases.
Further details of the model are given in \citet{2010A&A...519A.111B}.

\begin{figure}
\resizebox{\hsize}{!}{\includegraphics[bb=30 20 420 450,clip]{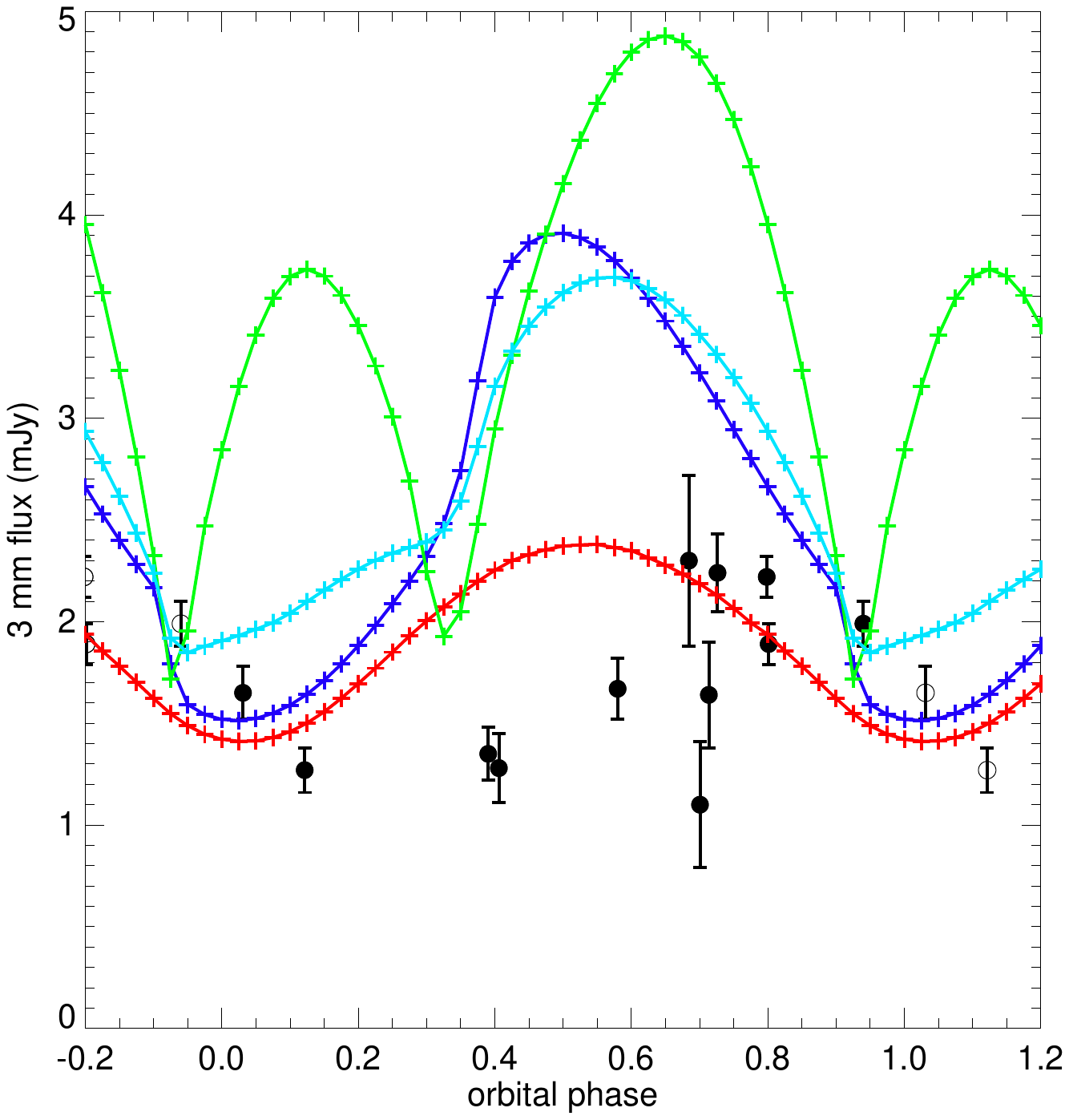}}
\caption{Theoretical fluxes compared to the observed 3~mm fluxes of 
\object{Cyg~OB2~\#8A}, plotted as a function of orbital
phase in the 21.908-day binary period. 
The black symbols with error bars show the observed data.
Open symbols indicate duplications in the extended phase range
(as in Fig.~\ref{fig fluxes}).
The dark blue curve is the synchrotron model, the red curve the
adiabatic thermal emission model, the green curve the radiative thermal
emission model, and the light blue curve the combined synchrotron
and adiabatic thermal emission model.
}
\label{fig theor fluxes}
\end{figure}

We applied the \citet{2010A&A...519A.111B} model
to \object{Cyg~OB2~\#8A}, using the 
stellar, stellar wind, and orbital parameters from
\citet[][see also Table~\ref{table parameters}]{2006MNRAS.371.1280D}. 
The simulation cube consists of 1024$^3$ cells, 
has a size of $4000~R_\sun$, and is centred on the primary. The stellar
wind material is assumed to be at 19\,000~K (i.e. about
half the effective temperature of the stars).

The resulting 3~mm fluxes are shown in Fig.~\ref{fig theor fluxes}
(dark blue curve). We note that the general flux level is too high,
as is the flux range. 
One should take into consideration that no parameters
were adjusted in the modelling, and some quantities are
not well known (e.g. the fraction of the shock energy that
is transferred to the relativistic electrons).
Because the predicted flux levels
coincide within an order of magnitude with the observations, the curve
can be viewed as indicative of the physical processes in play but it
clearly does not tell the whole story. 
Note that when 
\citet{2010A&A...519A.111B} applied this same model to the 6~cm observations
of \object{Cyg~OB2~\#8A}, they obtained theoretical fluxes that are 
systematically too low.
The maximum of the theoretical light curve is approximately as broad
as the observed one. The main discrepancy is that this theoretical maximum
is phase shifted with respect to the observations.

\citet{2010A&A...519A.111B} also presented a model with
stellar wind parameters for the primary that
are different from the \citet{2006MNRAS.371.1280D} ones
($\dot{M} = 1.0 \times 10^{-6}~{\rm M}_{\sun}{\rm yr}^{-1}$,
$\varv_\infty$ = 2500 ${\rm km\,s}^{-1}$). While this helped
to explain some features of the 6~cm observations, using these values
for the 3~mm observations only shifts the maximum even further
away from the observed one.

A possible explanation for the phase shift of the maximum
is our neglect of the orbital motion on the shape of the CWR.
On a large scale, the contact discontinuity and its associated shocks
take on the shape of a spiral, as shown by \citet{2008MNRAS.388.1047P}
using a simplified dynamical model. This spiral shape is confirmed
by the full hydrodynamical models of e.g. \object{$\eta$ Car} 
\citep{2011ApJ...726..105P,2013MNRAS.436.3820M}.
For the purposes of understanding the effect on the 3 mm fluxes
of \object{Cyg OB2~\#8A},
however, one needs to focus on a smaller scale, which extends
only somewhat beyond the $R_{\tau=1}$ distances of 220 and 140 $R_\sun$.
At this scale the orbital motion pushes the leading edge of the CWR
closer to the star with the weaker wind
\citep[see, e.g., Fig.~13 of][]{2011ApJ...726..105P}.
The position of the leading CWR edge in a model that does include orbital
motion is therefore approximately the same as the position at an earlier
phase in a model that does not include orbital motion. A first-order
correction of our theoretical light curve is therefore
a shift of the curve towards the right in the orbital phase diagram. 
This is indeed the correct
direction to improve the agreement of the theoretical and
observed phases of maximum. Of course, 
we cannot estimate the size of the
shift from this qualitative argument. For this we would need detailed
hydrodynamical calculations, which are outside the scope
of this paper.

\subsection{CWR thermal emission - adiabatic model}
\label{section adiabatic model}

In the previous section, we have only considered the contribution
of the synchrotron emission (and the stellar wind free-free emission
and absorption) to the observed fluxes.
It is expected that the highly compressed material in the CWR
will also contribute free-free emission and absorption, which could
be detectable in the observations
\citep{1995MNRAS.277..163S, 2010MNRAS.403.1633P, 2011A&A...531A..52M}.

To explore the effect of the thermal emission, 
we use another radiative transfer model, developed 
by \citet{2014A&A...561A..18B}. This model uses an
adaptive grid scheme allowing us to efficiently determine the emergent
intensities and flux from the stellar winds and the CWR. 
The CWR is assumed to have a shape that is
centred around a cone.
The position and opening angle of this cone are determined analytically
\citep[][their Eqs.~1 and 3]{1993ApJ...402..271E}. The CWR 
has a half flaring angle ($\alpha$), so it is thin at the apex
and expands as we move away from the line connecting the two stars
(see Fig.~2 of \citeauthor{2014A&A...561A..18B}).
The size of the CWR scales with the distance between
the two stars. We assume that the density inside the CWR is four times
higher than what the wind density would be at that distance
(this assumption will be revised in another model we discuss 
-- see Sect.~\ref{section radiative model}). 
We refer to the \citeauthor{2014A&A...561A..18B} paper for further details
of the model.

For each phase of the orbit, we put the two stars in a 
three-dimensional grid, and assign the densities and temperatures of
the stellar winds to each cell. The temperature
of the CWR material is a free parameter; the wind material is assumed to
be at half of the effective temperatures of the stars. 
The grid is $16\,000~R_\sun$ on each side, and we start with $256^3$ cells.
We then solve the radiative transfer equation
taking into account the free-free emission and absorption,
and refining the cells where needed (adaptive grid scheme).

We explored a range of values for the temperature of the CWR, its
flaring angle, and its size. The best fit we found (as judged by eye)
is shown on Fig.~\ref{fig theor fluxes} (red curve). 
Its parameters are a CWR temperature of $1.2 \times 10^5$~K,
a half-flaring angle of $30\degr$ and a size of
1.8 times the separation between the two stars.
These parameters were chosen from a large set of experiments,
because they get the average flux level and the
range of fluxes approximately correct. Similarly as for the synchrotron
model, however, the maximum is phase shifted with
respect to the observations. It is also broader than observed.
The use of the alternative \citet{2010A&A...519A.111B}
model (with a lower mass-loss rate for the
primary) does not solve the problem of the phase-shifted maximum.
Similarly as in Sect.~\ref{section CWR synchrotron emission},
the phase shift can probably be attributed to
our neglect of orbital motion.

To find out how much the CWR is contributing to the total flux,
we also ran a model without a CWR, by setting the flaring angle to
zero. We then find
that the 3 mm flux varies between 1.32 and 1.39 mJy, which is
just marginally below the minimum of the red curve in 
Fig.~\ref{fig theor fluxes}. The two stellar winds therefore provide
the ``baseline" flux, and the CWR is responsible for the variability
on top of that baseline flux.
The CWR region is undetectable around phase 0.0 -- 0.2 because the
primary with the strongest wind is in front of it (conjunction is
at phase 0.08). The free-free absorption in the stellar wind blocks
most of the flux from the CWR. 
Around phase 0.4, the two stars are
well separated on the sky, and that part of the CWR region which
is beyond the free-free absorption of the stellar winds becomes 
detectable. At later phases, the secondary with the weaker wind moves
in front (conjunction at phase 0.71), 
which reduces the detectable flux again, but not so
much as when the primary was in front.

\subsection{CWR thermal emission - radiative model}
\label{section radiative model}

The model in the previous section
limits the density contrast in the CWR to a factor
of four. This is appropriate for an adiabatic shock.
But, at least in part of the orbit the
shock will be radiative, and the density contrast could be higher. 
This can be seen from the values of
the cooling parameter $\chi$, as defined
by \citet{1992ApJ...386..265S}. 
\citet{2006MNRAS.371.1280D} derived $\chi$ values 
for \object{Cyg~OB2~\#8A} which are around one
(0.19 -- 1.65, depending on component and orbital phase; see their Table~11).
This indicates that the shocks are radiative at least part of the time
(near periastron, where $\chi \ll 1$), 
and can become adiabatic near apastron. 

We therefore adapted the \citet{2014A&A...561A..18B} code to include
the semi-analytical model for radiative shocks in colliding
stellar winds developed by \citet{2011A&A...531A..52M}.
This model assumes that the post-shock cooling is very efficient,
leading to a thin CWR with a
high density contrast, and a temperature that is equal
to the wind temperature.
\citeauthor{2011A&A...531A..52M} determine the position of the contact 
discontinuity semi-analytically, and from the conservation equations they 
derive the surface density and emission measure of the CWR material.
To avoid substantial changes to our code, we do not directly use the
emission measure, but we convert it into a density in
a region around the contact discontinuity. As a standard value, we take
20 $R_\sun$ for the width of that region.

\begin{figure}
\resizebox{\hsize}{!}{\includegraphics[bb=56 56 481 481,clip]{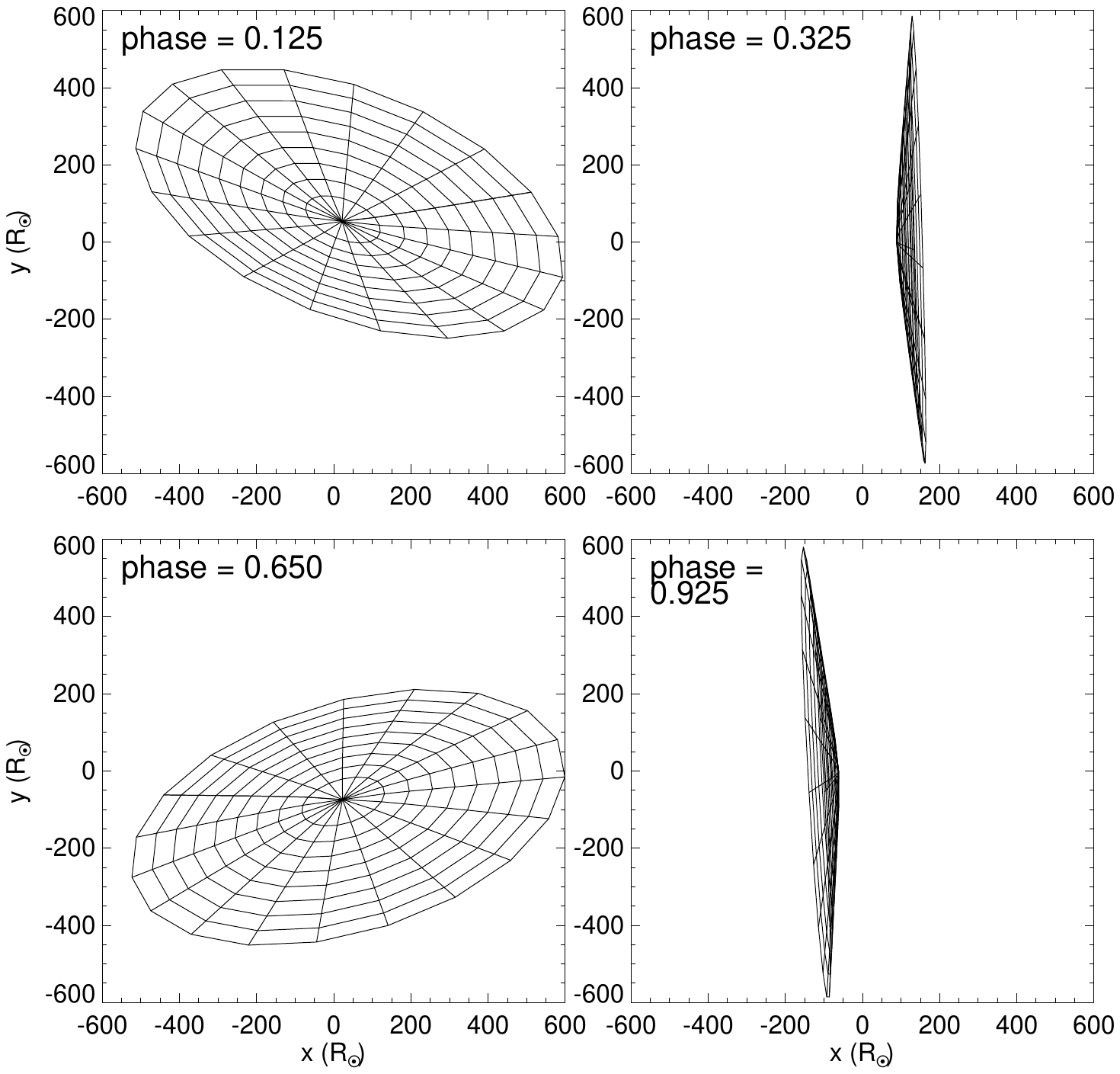}}
\caption{The CWR of the thin radiative shock model projected
on to the sky, for phases where the flux is high (left column)
and where the flux is low (right column). The orbital phase is listed
in the top left corner of each panel. For plotting purposes,
the CWR has been cut off at a distance of $600~R_\sun$.
}
\label{fig doublewave}
\end{figure}

Applying this model we get the results presented by the green 
curve on Fig.~\ref{fig theor fluxes}. The theoretical light curve
shows a clear double-wave pattern, which is not seen in the 
data. The double wave is explained in Fig.~\ref{fig doublewave},
where we plot the CWR projected against the sky for a few different
phases. When we see the CWR turned towards us (phases 0.125 and 0.65), 
there is a larger emitting area than when we see it on its side
(phases 0.325 and 0.925). This explains the minima and maxima
of the theoretical curve. 

An important element in this explanation is
that the CWR still has a significant flux contribution beyond the
free-free absorption region of the stellar winds. 
E.g., at phase 0.325, the separation of the two stars on the sky
is about $160~R_{\sun}$, which should be compared to the
$R_{\tau=1}$ radii of $220~R_{\sun}$ and $140~R_{\sun}$ 
(Table~\ref{table parameters}). Although these $R_{\tau=1}$ radii 
are based on an assumed spherically symmetric wind, they still
indicate that the part of the CWR closest its apex will not be detectable.
In contrast to this, for the adiabatic model from 
Sect.~\ref{section adiabatic model}
(red curve on Fig.~\ref{fig theor fluxes}) we have limited
the size of the CWR to 1.8 times the separation between the two stars,
so the CWR is only just beyond the $R_{\tau=1}$ radius.
Hence, it does not show the double wave effect.
Of course, also in the adiabatic model we could take a much larger
limit on the size of the CWR, and in that case we would also get a 
double wave pattern. As this does not fit the observations,
however, we did not present such models.
Applying the \citet{2010A&A...519A.111B} alternative stellar wind parameters 
(lower mass-loss rate for the primary) in the radiative model results in
a similar double-wave pattern.

\subsection{CWR combined synchrotron and adiabatic model}
\label{section combined model}

We can now extend the synchrotron model of 
Sect.~\ref{section CWR synchrotron emission}
to also include the free-free
emission and absorption of the CWR. In the \citet{2010A&A...519A.111B} computer code, we 
find those cells where the synchrotron emission is non-zero,
and we add the free-free opacity and emissivity, using a density
four times higher than the local stellar wind density. We consider
the temperature of the CWR wind to be a free parameter.

The resulting 3~mm fluxes are shown in Fig.~\ref{fig theor fluxes}
(light blue curve). We used a CWR temperature of $7.0 \times 10^4$~K,
which is lower than the value used in Sect.~\ref{section adiabatic model}
to avoid having fluxes that are too high. The combined model is comparable
in quality of fit to the pure synchrotron one. The phase of maximum 
is shifted slightly toward the observed maximum, but is still quite
some distance from it.

\subsection{Comparison to \citet{2010MNRAS.403.1633P}}

\citet{2010MNRAS.403.1633P} used his 3D hydrodynamical models of colliding-wind
binaries to predict the {\em thermal} emission from submillimetre to 
centimetre radio wavelengths. Three of the four models he calculated
have circular orbits, and are therefore not applicable to the
eccentric binary \object{Cyg~OB2~\#8A}. This is further confirmed
by the fact that all three predicted 3 mm light curves show a double
peak, which is not seen in the \object{Cyg~OB2~\#8A} data.

The \citet{2010MNRAS.403.1633P} cwb4 model is an eccentric binary, though with parameters
different from \object{Cyg~OB2~\#8A}: it is an O6V + O6V system
with a period of 6.1 days, and an eccentricity of $e=0.\overline{36}$.
Each of the components has a mass loss rate of
$\dot{M} = 2 \times 10^{-7}~{\rm M}_{\sun}{\rm yr}^{-1}$.
and a terminal velocity of $\varv_\infty$ = 2500 ${\rm km\,s}^{-1}$.
Nevertheless, the hydrodynamics are somewhat similar to what we
expect for \object{Cyg~OB2~\#8A}, in that the CWR is radiative at
periastron and adiabatic at apastron. A qualitative comparison
is therefore possible.

The cwb4 model shows a single peak in the 3 mm light curve, that
is close to periastron. The exact orbital phase of the peak
is only slightly dependent on the viewing angle. The observed data
however, peak around phase 0.8. The difference could be due to the fact
that the cwb4 model is for two equal winds, while the \object{Cyg~OB2~\#8A}
winds are not equal. This will shift the phase of maximum away
from periastron, but detailed hydrodynamical calculations of the
\object{Cyg~OB2~\#8A} system would be required to see if the maximum
moves to the observed one.
The flux level of the cwb4 peak is about three times the baseline flux.
This compares well to the factor of about two seen in the \object{Cyg~OB2~\#8A}
data.

\section{Cyg~OB2~\#8B}
\label{section 8B}

On the map of our combined data (Fig.~\ref{fig deep image}) we also detected 
\object{Cyg~OB2~\#8B} (= Schulte 8B = MT 462).
We measured its flux on the combined visibility data, by
fitting a model with two point sources (one for \#8A, and one for
\#8B). We found
a flux for \object{Cyg~OB2~\#8B} of 0.21 $\pm$ 0.033~mJy, with a
position that is offset by only 0\farcs4 from
its SIMBAD position.

In order to convert the observed flux into a mass-loss rate, we need to know
additional information about this star. Its spectral type
is listed as O6 II(f) \citep{2011ApJS..193...24S},
O6.5 III(f) \citep{1991AJ....101.1408M}, or
O7 III-II \citep{2007ApJ...664.1102K}.
There are no significant radial velocity changes
\citep{2014ApJS..213...34K}, making it unlikely that it is a binary.

There are no stellar parameter determinations for this star,
so we have to rely on calibrations to determine them. We follow
\citet{2016MNRAS.463..763M}, who use the \citet{2005A&A...436.1049M}
calibration to derive an effective temperature $T_{\rm eff} = 35644$~K, 
a luminosity of 
$\log L/L_\sun = 5.49$, and a (spectroscopic) mass of
$M/M_\sun = 33.68$,
based on the O6.5 III(f)
spectral type. From the 
\cite{1990ApJ...361..607P} calibration of terminal velocity versus
spectral type, $\varv_\infty$ = 2545 ${\rm km\,s}^{-1}$ follows.
Using this in Eq.~\ref{equation Wright Barlow}, we find
$\dot{M} = 1.42 \pm 0.2 \times 10^{-6}~{\rm M}_{\sun}{\rm yr}^{-1}$.

We can compare this to the upper limits that have been 
derived from radio observations.
\citet{2016MNRAS.463..763M} derive an upper limit 
from their 21 cm observations of 0.078~mJy, corresponding
to $\dot{M} < 4.3 \times 10^{-6}~{\rm M}_{\sun}{\rm yr}^{-1}$.
The 0.2~mJy upper limit at 6~cm found by \citet{1989ApJ...340..518B}
leads to
$\dot{M} < 5.2 \times 10^{-6}~{\rm M}_{\sun}{\rm yr}^{-1}$.
There are no determinations from other mass loss indicators
(such as H$\alpha$).
It should be noted that all mass-loss rates listed here
have not been clumping corrected.

The predicted mass-loss rate according to the \citet{2001A&A...369..574V}
formulae is $0.7 \times 10^{-6}~{\rm M}_{\sun}{\rm yr}^{-1}$. The
difference between this value and the observed one can be interpreted
as due to clumping, with a clumping factor of four. This value is
compatible with current ideas about clumping in massive-star
winds \citep[e.g.,][]{2008A&ARv..16..209P}.

\section{Conclusions}
\label{section conclusions}

We monitored the massive colliding-wind binary \object{Cyg~OB2~\#8A} 
at 3~mm with
the NOEMA interferometer, with good phase coverage of its orbit.
For 12 of the 14 observations, we could determine the flux.

The 3~mm light curve shows clear phase-locked variability, indicating that
a substantial part of the flux comes from the colliding-wind region 
(CWR)\footnote{One could consider alternative explanations, such as 
  phase-locked variability of the mass-loss rate, possibly caused by
  changing radiative inhibition in this eccentric binary
  \citep{1994MNRAS.269..226S}. However, also in that case a colliding-wind
  region would of necessity exist. The dominant effect of the CWR is
  shown by the fact that our models of the CWR can clearly get the
  magnitude of the flux variations correct -- though many important details
  still remain to be solved.
}.
We modelled the data using models that include synchrotron
radiation due to the CWR, or free-free emission from the CWR, or both.
All models give fluxes and flux ranges higher than those observed,
though comparable in magnitude to the observations. With the exception
of the radiative shock model, all have a single peak -- as observed.
The slightly better agreement of the
shape of the maximum
would favour the synchrotron interpretation.
From a modelling point of view, however, some contribution
of free-free emission from the CWR is also expected.
In summary, stronger arguments will be required for a decision
on how much each of the two emission mechanisms
contribute to the observed fluxes.

The main problem with the models presented here is that the 
theoretical maxima are phase shifted with respect
to the observed one.
It is clear that, at least qualitatively, this is due to our neglect of
the orbital motion on the shape of the CWR. 
To see if this is also quantitatively the correct explanation,
more sophisticated modelling will
be required. This should be based on solving the hydrodynamical equations,
in order to derive the density and temperature distribution in 
the CWR. These can then be used to calculate a better theoretical
light curve and will allow us to determine the relative contributions
of the synchrotron and free-free radiation emission. 

On the deep 3~mm image made by combining all our data, we also
detected the visual companion
\object{Cyg~OB2~\#8B}. We measured a flux of $0.21 \pm 0.033$~mJy,
which leads to a (non-clumped) mass-loss rate of
$\dot{M} = 1.42 \pm 0.2 \times 10^{-6}~{\rm M}_{\sun}{\rm yr}^{-1}$.
From a comparison with predicted mass-loss rates,
we derive a clumping factor of four, which is compatible with current
ideas about clumping in massive-star winds.

\begin{acknowledgements}
We are especially grateful to Jan Martin Winters and \mbox{Charl\`ene}
Lef\`evre (IRAM) for their assistance with the data reduction.
We thank the referee for his/her comments, which helped to improve the paper.
We have made use of the GILDAS software
for the reduction of the data.
This research has made use of the SIMBAD database,
operated at CDS, Strasbourg, France and NASA's
Astrophysics Data System Abstract Service.
J. Morford
and D. Fenech wish to acknowledge funding from an STFC studentship and
STFC consolidated grant (ST/M001334/1) respectively.
\end{acknowledgements}

\bibliographystyle{aa}
\bibliography{31403}

\end{document}